\documentclass[12pt,preprint]{aastex}

\def\beq{\begin{equation}}
\def\eeq{\end{equation}}
\def\bey{\begin{eqnarray}}
\def\eey{\end{eqnarray}}
\def\kms{\mbox{\rm \,km\,s}^{-1}}
\def\kpc{\rm kpc}
\def\rel{{\rm rel}}
\def\msun{M_\odot}
\def\thalf{t_{1/2}}

\def\te{t_{\rm E}}
\def\e{{\rm E}}

\def\thetae{\theta_{\rm E}}
\def\max{{\rm max}}
\def\min{{\rm min}}

\def\solmasspc{{\msun {\rm pc}{}^{-2}}}
\def\l{{\rm L}}
\def\s{{\rm S}}
\def\ls{{\rm LS}}
\def\bv{{\bf v}}
\def\spose#1{\hbox to 0pt{#1\hss}}
\def\lta{\mathrel{\spose{\lower 3pt\hbox{$\sim$}}
    \raise 2.0pt\hbox{$<$}}}
\def\gta{\mathrel{\spose{\lower 3pt\hbox{$\sim$}}
    \raise 2.0pt\hbox{$>$}}}

\begin{document}

\title{A Candidate M31/M32 Intergalactic Microlensing Event}

\author{S.~Paulin-Henriksson$^1$, 
P.~Baillon$^2$, A.~Bouquet$^1$, B.J.~Carr$^3$, 
M.~Cr{\'e}z{\'e}$^{1,4}$, N.W.~Evans$^5$, Y.~Giraud-H{\'e}raud$^1$, 
A.~Gould$^{1,6}$, 
P.~Hewett$^7$, J. Kaplan$^1$, E.~Kerins$^8$, 
Y.~Le~Du$^{1,5}$, A.-L.~Melchior$^{9}$,
S.J.~Smartt$^7$ and D.~Valls-Gabaud$^{10}$\\ 
\smallskip
\centerline{(The POINT--AGAPE Collaboration)}}

\affil{
$^1$Laboratoire de Physique Corpusculaire et Cosmologie, 
Coll{\`e}ge de France, 11 Place Marcelin Berthelot, F-75231 Paris, France\\
$^2$CERN, 1211 Gen{\`e}ve, Switzerland \\
$^3$Astronomy Unit, School of Mathematical Sciences, Queen Mary,
    University of London, Mile End Road, London E1 4NS, UK\\
$^{4}$Universit\'e Bretagne-Sud, campus de Tohannic, BP 573, F-56017 
    Vannes Cedex, France\\
$^5$Theoretical Physics, 1 Keble Road, Oxford OX1 3NP, UK\\
$^6$Department of Astronomy, Ohio State University, 140 West 18th Avenue, Columbus, OH 43210\\
$^7$Institute of Astronomy, Madingley Road, Cambridge CB3 0HA, UK\\
$^8$Astrophysics Research Institute,
Liverpool John Moores University, 12 Quays House, Egerton Wharf, Birkenhead,
CH41 1LD, UK\\
$^{9}$LERMA (Laboratoire d'Etudes du Rayonnement
et de la Matière en Astrophysique), FRE2460, Observatoire de Paris,
61, avenue de l'Observatoire, F-75014 Paris, France.
$^{10}$Laboratoire d'Astrophysique UMR~CNRS~5572, Observatoire
    Midi-Pyr{\'e}n{\'e}es, 14 Avenue Edouard Belin, F-31400 Toulouse,
    France}

\begin{abstract}
We report the discovery of a microlensing candidate projected $2'54''$
from the center of M32, on the side closest to M31.  The blue color
$(R-I= 0.00\pm 0.14)$ of the source argues strongly that it lies in
the disk of M31, while the proximity of the line of sight to M32
implies that this galaxy is the most likely host of the lens.  
If this
interpretation is correct, it would confirm previous  arguments that M32 lies 
in front of M31.  We estimate that of order one such event or less should be
present in the POINT-AGAPE data base.  If more events are discovered
in this direction in a dedicated experiment, they could be used to measure 
the mass function of
M32 up to an unknown scale factor. By combining microlensing
observations of a binary-lens event with a measurement of the M31-M32
relative proper motion using the astrometric satellites {\it SIM} or
{\it GAIA}, it will be possible to measure the physical separation of
M31 and M32, the last of the six phase-space coordinates needed to
assign M32 an orbit.
\end{abstract}


\section{{Introduction}
\label{sec:intro}}

Following the suggestions of \citet{crotts92} and \citet{baillon93},
the POINT-AGAPE collaboration\footnote{ see {\tt
http://www.point-agape.org}} is carrying out a pixel lensing survey of
M31 using the Wide Field Camera (WFC) on the Isaac Newton Telescope
(INT).  We monitor two fields of 0.3 deg$^2$ each, located North and
South of the M31 center.  The main goal of the survey is to map the
global distribution of microlensing events in M31 and to determine any
large-scale gradient. M31 is highly inclined, so there will be a
strong gradient if a substantial fraction of the lenses lie in the
dark halo of M31.  The main difficulties are that the M31 sources are
resolved only while they are lensed (and then only if the
magnification is substantial), and seeing causes substantial
variations in the point spread functions. The pixel lensing technique
has been developed to cope with these problems \citep{ans97,ans99}.

        The large field of view happens to encompass M32, the
dwarf elliptical, whose center lies $25'$ from the center of
M31, corresponding to $\sim 5$ kpc in projected distance.  Since M32
lies $\sim 53.\hskip-2pt ^\circ 5$ from the far-minor axis of M31, the M31 disk
stars at this projected position are $\sim 10$ kpc from the center of
M31.  The M31 disk at this position is relatively faint and blue.   Since
M32 probably lies in front of M31 \citep{fjj}, this opens up the 
possibility of detecting
microlensing events of M31 stars by M32 stars (or other compact
objects).  Here, we report such a candidate M31/M32 intergalactic
microlensing event.  We argue from the geometry and color of the event
that the intergalactic interpretation is the most plausible.  The
discovery of additional events of this type could be used determine
the M32 mass function, as well as the relative distances and
velocities of M31 and M32.

\section{{Observations and Data Analysis}
\label{sec:obsdata}}

Observations of the event, named PA-00-S4, were obtained in $r'$ and
$i'$ bands (similar to Sloan $r'$ and $i'$) from 2000 August to 2001
January. The field was also observed in $g'$ and $r'$ bands from 1999
August to 2000 January, when the source was quiescent.  Total exposure
times were between 5 and 10 minutes per night.  Data reduction is
described in detail elsewhere \citep{ans97,ledu00}.  After bias
subtraction and flat-fielding, each image is geometrically and
photometrically aligned relative to a reference image (1999 August 8),
which was chosen because it has a long exposure time, typical
\mbox{seeing ($1.\hskip-2pt ''5$)} and little contamination from the
Moon.  The lightcurves are computed by summing the flux in 7-pixel
($2.\hskip-2pt ''3$) square ``superpixels'' and removing the
correlation with seeing variation.  The transformation from
instrumental $(r',i')$ to Cousins $(R,I)$ is based on $\sim 50$
standards (\citealt{haiman}; catalogue II/208 of VizieR) that lie on
the same CCD as PA-00-S4.

We use a simple set of criteria to select candidates from two seasons
of INT WFC data.  Detection of events is made in
the $r'$ band, which has better sampling and is free of fringing effects 
on the CCDs.  We fit all lightcurves having detectable bumps to
a \citet{Paczynski86} curve with seven parameters: the Einstein timescale
$t_\e$, the time of maximum, $t_0$, the impact parameter (in units of the
Einstein radius) $u_0$, and two flux parameters for each filter, one for
the source $F_s$, and one for the background $F_b$.  
A bump is defined by at least 3 consecutive $r'$ data points rising above the 
baseline by at least $3\sigma$, with at least 2 points 
 (in either band) on both the rising and
falling parts of the variation (defined as the interval over which
the lightcurve is at least $3\,\sigma$ above baseline). We 
calculate the probability $P$ that the bump is due to
random noise, and demand $-\ln P > 100$ in $r'$ and $-\ln P > 20$ in the
second filter ($i'$ for 2000).  
To allow for non-standard microlensing events, we 
initially set a loose threshold of $\chi^2/\rm dof <5$.  
Then, to extract a sample of high signal-to-noise ratio events, we demand 
$R(\Delta F) <21$, where $R(\Delta F)$ 
is the (Cousins) magnitude corresponding to 
$(A_\max-1) F_s$ and $A_\max$ is the peak of the Paczy\'nski fit.
After eliminating the lightcurves with strong secondary bumps, there
remain 362 candidates,  of which 8 have FWHM shorter than 25 days.
Four of these 8 short candidates are almost certainly microlensing events
\citep{fourev}.  Among them is PA-00-S4, with a projected position 
close to the center of M32.

\section{{The M31/M32 Candidate}
\label{sec:candidate}}

\begin{figure}
\plotone{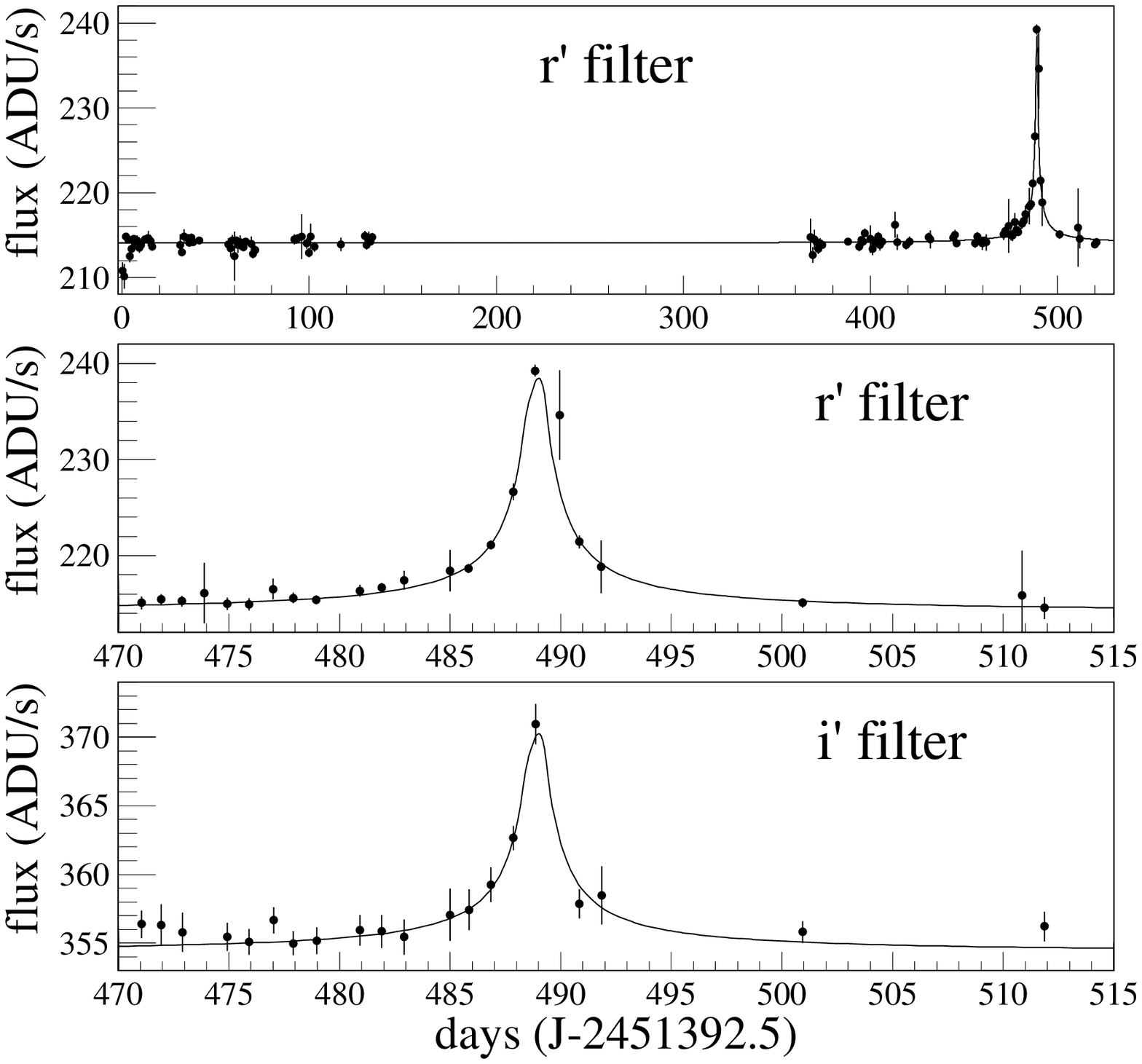}
\caption{\label{fig:lightcurves}
Lightcurves for the M31/M32 intergalactic candidate microlensing event
PA-00-S4.  Upper panel shows full two years of INT WFC data in $r'$.  Lower 
two panels show zooms of the event in $r'$ and $i'$.  Both are well fit
by a Paczy\'nski curve with a single set of geometrical parameters.
}\end{figure}

Figure \ref{fig:lightcurves} 
shows the lightcurves in $r'$ and $i'$ of PA-00-S4 together 
with the Paczy\'nski fit. The source has J2000 
position $\alpha=00^\textrm{h}42^\textrm{m}30.0^\textrm{s}$, 
$\delta=+40^\circ53'47.\hskip-2pt ''1$. 
That is, it lies projected on the far disk of 
M31, $22'31''$ from the center of M31, with a position angle $\sim 
59. \hskip-2pt ^\circ 2$ relative to the minor axis. It also lies $2'54''$ 
from the center of M32. There are some straightforwards tests to see if 
PA-00-S4 is compatible with microlensing. First there are no comparable 
``bumps'' in the remainder of the lightcurve, as might 
be expected for many classes of variable stars. Second, it is achromatic: 
the $r'$ and $i'$ data are simultaneously fit to a Paczy\'nski curve and
show no significant systematic offset relative to one another.
The FWHM of the peak is $\thalf=2.1\,$days, and at 
the maximum magnification the source is $R=20.7\pm 0.2$ (corresponding to
$M_R=-3.9$) and $R-I=0.00\pm 0.14$.  We know of no variables 
capable of producing a time-symmetric outburst of $2000\,L_\odot$ on
such a short timescale.  Note that the magnitude error is greater than
the color error because the former is more affected by seeing.  
We assume a ratio of selective-to-total exinction
$E(B-V)=0.062$ \citep{schlegel} and a distance of 770 kpc.

\section{{Lightcurve Analysis}
\label{sec:lightcurve}}

\subsection{{Source Location}
\label{sec:sourceloc}}

	Microlensing events of unresolved sources are generically subject 
to a degeneracy in which the product of source flux $F_s$ and the timescale 
$t_\e$ is much better determined than either parameter separately.        
PA-00-S4 suffers from this degeneracy at about the factor 2 level:
\begin{equation}
\label{eqn:terstar}
\log{t_\e\over\rm day}=2.11\pm 0.34,\qquad R_s = 26.65\pm 0.85.
\end{equation}
Given the magnitude at peak, $R=20.7$, these 
correspond to a maximum magnification
$A_\max=240^{+285}_{-130}$.  
At the distance and reddening of M31, the source flux corresponds to 
$M_R=2.05\pm 0.85$.
This absolute magnitude and the dereddened color, $(R-I)_0=-0.05\pm
0.13$, are consistent with the source being either an A type
main-sequence star or a blue horizontal branch (BHB) star.  The former
are expected to be common in the M31 disk, while about
8000 of the latter have been counted within a near-central $0.45\,\rm arcmin^2$
of M32.  Given the short evolutionary phase of the BHB and the fact that
the surface brightness of M31 is about twice that of M32 at this location,
the source is strongly favored to be an M31 A star, even prior to the
auxiliary information presented in \S~\ref{sec:lensloc}.

\subsection{{Lens Location}
\label{sec:lensloc}}

Current distance-indicator based estimates do not permit one
to say whether M31 or M32 is closer (e.g., \citealt{mateo}).  However,
if M32 lay behind M31, it would suffer extinction due to dust in
the M31 disk.  The absence of such extinction as determined from a wide
variety of observations led \citet{fjj} to conclude that M32 must be
in front.  Moreover, the disk of M31 is disturbed
\citep{argyle,einrum,gottdav}, leading
\citet{arp} and \citet{roberts} to suggest an encounter with M32 as
the cause.  Byrd's (1976) model of this encounter, which has M32
passing right through the disk of M31, places M32 today 8.5 kpc in front
of M31, i.e. $\sim 20\,\kpc$ in front of the M31 disk stars projected
along the line of sight.  In any case, M32 certainly lies in M31's
potential well and orbits it rather closely: \citet{faber} argued from
the profile of M32 that it had been tidally stripped by encounters
with M31, and this has now been proved by the discovery of a
Sagittarius-dwarf like tidal stream associated with M32 \citep{ibata}.

	For the moment, we assume that M32 lies 20 kpc in front of M31,
in accordance with the estimate of \citet{byrd}.  Then there are 
six
possible locations for the lens: the disk or halo of M31, the disk or
halo of the Galaxy, M32 itself, or the tidal stream associated
with it.  The optical depth for any
population, $i$, of lenses is $\tau_i=(4\pi G/c^2)\Sigma_i D_i$, where
$\Sigma_i$ is its surface mass density, and $D_i=
D_{\l,i}D_{\ls,i}/D_\s$, a combination of the distances between the
observer (O), lens (L), and source (S).  For the populations under
consideration, $D\simeq\min\{D_\l,D_\ls\}$.

The dereddened surface brightness of M32 at the position of the event is
$R=22.1\,\rm mag\,arcsec^{-2}$.  Adopting a stellar (actually,
compact-object) mass-to-light ratio of $M/L_R=3$, this implies a
surface density $\Sigma_{\rm M32}=110\,\solmasspc$.  The optical depth
is therefore $\tau\sim 1.4\times 10^{-6}(D_{\rm M31}-D_{\rm
M32})/20\,\kpc$.  This is one to several orders of magnitude larger than the
optical depths of the M32 tidal stream, 
the disks of M31 \citep{gould94} and the Milky Way \citep{gbf97}, and to the
optical depth due to compact objects in the Milky Way halo
\citep{macho,eros}.  The total optical depth of the M31 halo at this
location, if it were entirely composed of compact objects, would be
$\tau_{\rm M31}\sim 4\times 10^{-6}$.  However, our current data already
rule out a full MACHO halo for M31 \citep{fourev}.
If only 20\% of the M31 halo
were in compact objects, as \citet{macho} have argued is true
for the Milky Way, then the M31 halo optical depth would be about
half that due to stars in M32.  We conclude that the most
likely location of the lens is a star in M32 but that it could be
a compact object in the halo of M31 as well.

\subsection{{Plausibility}
\label{sec:plausible}}

        What is the probability of finding a microlensing event for
which the lens lies in M32 and the source in M31, and what is the most
likely projected separation between such an event and the center of
M32?  At any given time, the expected number of events per unit
surface area is
\begin{equation}
{d n\over dA} = {4\pi G \over c^2}N_\s(x,y)\Sigma_\l(x,y)D_\ls
\label{eqn:difprob}
\end{equation}
where $\Sigma_\l$ is the surface mass density of lenses and $N_\s$ is the
column density of sources.  Integrating equation (\ref{eqn:difprob}) over
the entire M32 galaxy, and making the assumption (true to first order)
that the source and lens densities are not correlated, yields
\begin{equation}
n_0 = {4\pi GM_{\rm M32}\over c^2}D_\ls\langle N_\s\rangle
= 40\,{M_{\rm M32}\over 10^9\,M_\odot}\,{D_\ls\over 20\,\kpc}\,
{\langle N_\s\rangle\over 3\,\rm pc^{-2}}.
\label{eqn:intprob}
\end{equation}
The quantities at the denominator of various ratios 
have been chosen to make those ratios close to unity. These adopted 
normalizations are derived as follows.  For the total mass
of M32, we assume an integrated luminosity of $M_B=-15.5$ \citep{kt},
a color of $B-V=0.9$ and a compact-object mass-to-light ratio of
$M/L_V=3.5$.  We measure the M31 dereddened surface brightness at the radial
position of M32 to be $R=21.4\,\rm mag\,arcsec^{-2}$.  If all of this
light were typically coming from $M_R=1$ stars, then there would be
3 pc$^{-2}$ such stars.  Only
relatively high magnification events would be detectable with our
current experimental setup.  The number taking place at any given time
 with $A_\max>A_{\max,\rm thresh}$ is $n \simeq n_0/A_{\max,\rm
thresh}$.  To meet our selection criterion $R(\Delta F)<21$ requires
$A_\max>70$. The measured color, $(R-I)_0\sim -0.05\pm 0.13$ and the
color-luminosity relation of main-sequence stars favor a source
luminosity at the bright end of the 
range given in equation (\ref{eqn:terstar}),
and so a timescale at the short end.  We adopt $t_\e\sim 60\,$days.
Since $\pi t_\e/2\sim 95\,$days, we would detect
$\sim (40/70)\times (150/95)\sim 1$ per observing season, assuming 100\%
efficiency.  While we have not yet calibrated our efficiency, we
expect that it is likely to be $\la 10\%$ for this type of event
$(A_\max>75$, $t_\e\sim 60\,\rm days)$ averaged over the relatively dense
field of M32 (see below).
So, we were somewhat lucky to detect such an event in the
POINT-AGAPE database for 1999 and 2000, but not excessively so.  

\begin{figure}
\plotone{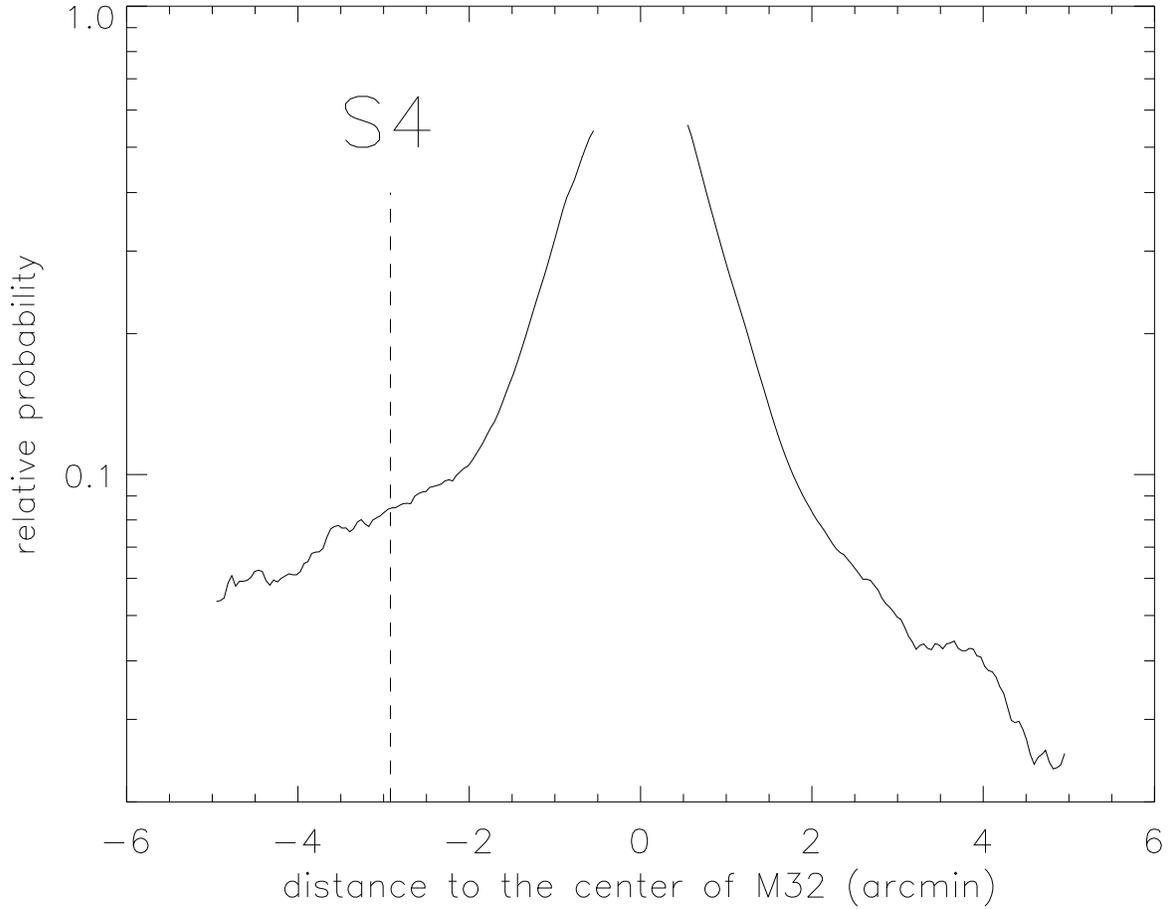}
\caption{\label{fig:relativelike}
Relative probability for an event to occur in semi-circular radial bins
around M32 on the sides closer to (negative) and farther from (positive) 
M31.  The curve does not go to the center of M31, which is saturated.
The event PA-00-S4 lies on the shoulder of this distribution, about a
factor of 10 below the peak.  However, the probability of {\it detecting}
an event is roughly flat as a function of radius because the higher
surface brightness interferes with detection.
}\end{figure}

        At what projected separation are we most likely {\it to detect} such
an event?  The surface brightness of M31 falls rapidly across the face
of M32.  We therefore first evaluate the probability for an event 
{\it to occur} in semicircular radial bins on each side 
of M32 by integrating the product of the M31 and M32 surface brightnesses
over the bin (see eq.\ [\ref{eqn:difprob}]).
Figure \ref{fig:relativelike} shows the result.  As one moves from the center
of M32 closer to M31, the probability is enhanced by the higher
density of M31 sources, and by the larger size of the semicircle, but
is degraded by the falling surface density of M32.  On the side that
is farther from M31, the surface brightness of both galaxies falls off
rapidly, leading to a rapid decline in probability.  Near the very
center, our images are saturated, so there is no possibility of
detection.  The event PA-00-S4 lies on the shoulder of the probability curve,
about a factor 10 below the peak.  However, while events {\it occur} much
more frequently near the center, they are not much more likely to be {\it
detected}.  Specifically, we find that a similar plot of variable stars 
detected with $R(\Delta F)<21$ is roughly flat from the position of PA-00-S4
to the cutoff at $\sim 0.\hskip-2pt '5$.  We conclude that the position of
PA-00-S4 is a plausible place to detect such an event.

\section{{Possible Applications}
\label{sec:application}}

        There is a limited amount that can be concluded from one
event, especially as we cannot be certain that the lens
resides in M32 (see \S\ \ref{sec:lensloc}).  Nevertheless, it is worth
considering the scientific applications of an ensemble of microlensing
events in the neighborhood of M32, gathered not just in the
experiments monitoring M31 but also perhaps in a future survey
centered on M32 itself.  First, if the events were clustered on the
M31 side of M32 as predicted by Figure \ref{fig:relativelike}, then
this would prove incontrovertibly that M32 lies in the foreground of
the M31 disk.

        Second, the distribution of Einstein timescales $\te$ can be
used to measure the mass function of M32, modulo an initially unknown
scale factor.  The timescale is given by $\te=\thetae/\mu_\rel$, where
$\thetae$ is the angular Einstein radius and $\mu_\rel$ is the
source-lens relative proper motion.  When, as in the present case,
$D_\ls\ll D_\s$, this reduces to
\begin{equation}
\te \sim
{1\over v_{\rel,\perp}}\sqrt{4 G M D_\ls \over c^2} = 73\,{\rm days}
\biggl({v_{\rel,\perp}\over 300\,\kms}\biggr)^{-1}
\biggl({M_l \over M_\odot}\biggr)^{1/2}
\biggl({D_\ls \over 20\,\kpc}\biggr)^{1/2},
\label{eqn:tedef}
\end{equation}
where $v_{\rel,\perp} = |\bv_{\l,\perp} - \bv_{\s,\perp}|$ is the
lens-source relative transverse velocity.  The transverse velocity of
the M31 disk at this location is known from its rotation curve and
inclination and is $\sim 200\,\kms$.  The transverse velocity of M32
is unknown, but since M32 is in the potential of M31, its magnitude is
likely to be $\sim 200\,\kms$.  Since these speeds are much larger
than the internal dispersions of either population, $v_{\rel,\perp}$
is probably $\sim 300\,\kms$ and, very importantly, roughly the same
for all events.  Thus, if the timescales can be measured, the masses
can also be measured, up to the unknown scale factor
$v_{\rel,\perp}^2/D_\ls$.  We are unable to measure the timescale
precisely in the case of PA-00-S4 because the source is not resolved
in ground-based images, and so we do not know the source brightness
which is degenerate with $\te$ (see eq.\ [\ref{eqn:terstar}]).  However,
M31 microlensing sources can be resolved with the {\it Hubble Space
Telescope} (e.g., \citealt{gold}). Indeed, virtually all potential
sources in the field of M32 could easily be resolved with just 4
snapshots using the Advanced Camera System.  Note that this
mass-function measurement is very similar to Paczy\'nski's (1994)
proposal to measure the mass function of globular clusters that lie
projected against the bulge, except that the distances and proper
motions of the clusters are already reasonably well known, so there is
no unknown scale factor.

        Third, the ensemble of events would yield the product $M_{\rm
M32} D_\ls$, according to equation (\ref{eqn:intprob}).  Combined with
an independent estimate of $M_{\rm M32}$ (say, by assuming a
compact-object mass-to-light ratio or by stellar dynamical modelling),
this would give an estimate of $D_\ls$.

        Fourth, if a caustic-crossing binary-lens event were observed,
one could use it to measure $v_{\rel,\perp}$.  That is, if the source
can be resolved in two bands, its surface brightness can be inferred
from its color, and hence its angular size can be determined from its
flux.  The time that the source takes to cross the caustic (plus the
crossing angle, which can be inferred from the solution of binary-lens
geometry), then yields $\mu_\rel=v_{\rel,\perp}/D_\s$.  This technique
has been successfully applied to five events
\citep{mb28,mb41,ob23,smc,eb5}.  A determination of $v_{\rel,\perp}$
would remove some of the degeneracy in the mass-function measurement.
This requires that the event be observed during the caustic crossing,
which only lasts a time $2\Delta t = 2r_*/v_{\rel,\perp} \csc\phi=
4\,{\rm hr}(r_*/3r_\odot)/(v_{\rel,\perp}/300\,\kms)\csc\phi$, where
$\phi$ is angle between the source-lens relative motion and the normal
to the caustic.  Hence, the crossing is unlikely to be observed unless
special preparations are taken. Of course, this requires predicting
the crossing in real time, as was done in all previous cases to which
this technique has been applied.

        Finally, we look ahead to combining microlensing and
space-based interferometry to obtain all six phase-space coordinates
of M32 relative to M31.  At present, three of these coordinates are
known (the two transverse position coordinates and the relative radial
velocity), while the remaining three are unknown. The proper motion of
M32 is $\mu_{\rm M32} = 60\mu\rm as\,yr^{-1}(v_{\l,\perp}/200\,\kms)$.
It will easily be measured by the {\it Space Interferometry Mission
(SIM)} with its $\sim 4\,\mu$as precision and 5-yr baseline. It will
also be measured by the scanning satellite {\it GAIA}, which will
record the proper motions of $\sim 10^4$ stars in M32 with $\sim
100\,\mu$as precision, from which the proper motion of M32 itself is
obtainable with $\sim 1\,\mu$as precision.  Once the proper motion is
determined, not only the magnitude of $\bv_{\rel,\perp}$, but also its
direction will be known.  As we now describe, this can be combined
with microlensing parallax information to determine $D_\ls$.

        Normally, one does not think of parallax effects in relation to
M31 microlensing, because the projected Einstein radius is so large,
\begin{equation}
\tilde r_\e = \sqrt{4 G M D_\s D_\l\over D_\ls c^2} \sim
460\,{\rm AU}\sqrt{M/M_\odot\over D_\ls/20\,\kpc},
\label{eqn:retilde}
\end{equation}
that it dwarfs any conceivable parallax baseline, whether space-based
\citep{refsdal} or ground-based \citep{gould92}.  However, parallax
effects on a caustic crossing scale inversely with the projected source
radius
\begin{equation}
\tilde r_* = {D_\l\over D_\ls}r_* \sim
0.5\,{\rm AU}\biggl({D_\ls\over 20\,\kpc}\biggr)^{-1} {r_*\over 3 r_\odot}
\label{eqn:rstartilde}
\end{equation}
rather than as $\tilde r_\e^{-1}$ \citep{hw,ga}, and therefore could
be precisely measured by a satellite in an Earth-like orbit even with
relatively modest photometric capabilities.  The time delay
between the caustic crossings as seen from the Earth and satellite is
\begin{equation}
\delta t = {D_{\rm sat}D_\ls\cos(\gamma-\phi)\over D_\s
D_\l\mu_\rel\sin\phi},
\label{eqn:deltat}
\end{equation}
where $D_{\rm sat}$ is the magnitude of the Earth-satellite separation
vector projected onto the plane of the sky, $\gamma$ is the angle
between this projected vector and direction of source-lens relative
motion, and where $\phi$ is again the angle between this motion and
the normal to the caustic.  As discussed above, $\mu_\rel$ can usually
be measured from caustic crossings and $\phi$ can be determined from
the overall solution to the binary event.  Unfortunately, $\gamma$ is
not generally known, and hence the measurement of $\delta t$ would not
typically lead to a determination of $D_\ls$.  However, since {\it
SIM} or {\it GAIA} will measure the {\it vector} proper motions of M31
and M32
(and so determine the orientation of the M31-M32 relative proper
motion as well), $\gamma$ will in fact be known.

\section{Conclusions}

We have reported the discovery of a high S/N, short-duration
microlensing candidate, PA-00-S4. It is a remarkable event
because the source almost certainly lies in the M31 disk, yet the lens
most probably resides in M32. This makes it the first convincing
candidate for an M31/M32 intergalactic microlensing event.  If this
interpretation is correct, it demonstrates that M32 lies in front of
M31.  We estimate that $\la 1$ such event satisfying the selection
criterior $R(\Delta F)<21$ should be
detectable in the current POINT-AGAPE microlensing survey of M31.

The scientific applications of gathering an assemblage of events
centered around M32 are substantial. If the events are clustered on
the M31 side, this provides unambiguous evidence that M32
lies in front of M31. The events can be used to deduce the mass
function of M32 up to an unknown scale factor.  Microlensing
observations of a binary-lens event, together with a measurement of
the M31-M32 relative proper motion using astrometric satellites like
{\it SIM} or {\it GAIA}, will enable the line-of-sight separation of
M31 and M32 to be determined.  This provides strong motivation for a
microlensing survey targetted on M32 itself.

\bigskip
\noindent
{\bf Acknowledgments}: YLD is supported by a PPARC postdoctoral
fellowship and SJS by a PPARC advanced fellowship. NWE acknowledges
help from the Royal Society. Work by AG was supported in part by a
grant from the Centre National de Recherche Scientific and in part by
grant AST 97-27520 from the NSF.

\end{document}